\documentclass[twocolumn,pre,showpacs,superscriptaddress]{revtex4}

\usepackage{graphicx}
\usepackage{multirow}

\begin{document}

\newcounter{footnote1}
\newcounter{footnote5}
\newcounter{footnote61}
\newcounter{footnote8}

\title{
  Liquid state properties from first principles DFT calculations: Static
  properties
}


\preprint{LA-UR 10-02740}

\author{Nicolas Bock}
\email{nbock@lanl.gov}
\affiliation{Theoretical Division, Los Alamos National Laboratory, Los Alamos, NM, 87545}

\author{Erik Holmstr\"{o}m}
\affiliation{Instituto de F\'{\i}sica, Faculdad de Ciencias, Universidad
Austral de Chile, Casilla 567, Valdivia, Chile}
\affiliation{Theoretical Division, Los Alamos National Laboratory, Los Alamos, NM, 87545}

\author{Travis B.~Peery}
\affiliation{Theoretical Division, Los Alamos National Laboratory, Los Alamos, NM, 87545}

\author{Raquel Liz\'{a}rraga}
\affiliation{Instituto de F\'{\i}sica, Faculdad de Ciencias, Universidad
Austral de Chile, Casilla 567, Valdivia, Chile}
\affiliation{Theoretical Division, Los Alamos National Laboratory, Los Alamos, NM, 87545}

\author{Eric D.~Chisolm}
\affiliation{Theoretical Division, Los Alamos National Laboratory, Los Alamos, NM, 87545}

\author{Giulia De~Lorenzi--Venneri}
\affiliation{Theoretical Division, Los Alamos National Laboratory, Los Alamos, NM, 87545}

\author{Duane C.~Wallace}
\affiliation{Theoretical Division, Los Alamos National Laboratory, Los Alamos, NM, 87545}

\date{\today}

\begin{abstract}

  In order to test the vibration-transit (V-T) theory of liquid dynamics,
  \emph{ab initio} density functional theory (DFT) calculations of
  thermodynamic properties of Na and Cu are performed and compared with
  experimental data. The calculations are done for the crystal at $T = 0$ and
  $T_{m}$, and for the liquid at $T_{m}$. The key theoretical quantities for
  crystal and liquid are the structural potential and the dynamical matrix,
  both as functions of volume. The theoretical equations are presented, as well
  as details of the DFT computations. The properties compared with experiment
  are the equilibrium volume, the isothermal bulk modulus, the internal energy
  and the entropy.  The agreement of theory with experiment is uniformly good.
  Our primary conclusion is that the application of DFT to V-T theory is
  feasible, and the resulting liquid calculations achieve the same level of
  accuracy as does \emph{ab initio} lattice dynamics for crystals. Moreover,
  given the well established reliability of DFT, the present results provide a
  significant confirmation of V-T theory itself.

\end{abstract}

\maketitle

\section{Introduction}

The goal of our work is to investigate, and to improve where possible, the
theoretical procedures for calculating statistical mechanical properties of
condensed matter systems. Here we shall focus on elemental crystals and
liquids. For many such systems, \emph{ab initio} density functional theory
(DFT) provides highly accurate results for the groundstate energy as function
of the nuclear positions. This energy is the groundstate adiabatic potential,
which appears in the nuclear motion Hamiltonian. For crystals, the nuclear
motion Hamiltonian is prescribed by lattice dynamics theory \cite{BornHuang}.
Thermodynamic properties of elemental crystals, as calculated from DFT
together with phonon statistical mechanics, can be nearly as accurate as the
experimental measurements. For liquids, in the absence of a tractable nuclear
motion Hamiltonian, statistical mechanical properties are calculated from
\emph{ab initio} molecular dynamics (MD). These calculations are based on DFT,
and can be evaluated in the groundstate adiabatic approximation
\cite{RevModPhys.64.1045, PhysRevB.47.558}.  Again the results compare very
well with experiment. Moreover, the studies reveal detailed characteristics of
the electronic structure, for example for liquid Al \cite{PhysRevB.57.8223},
for liquid Fe \cite{PhysRevB.61.132, PhysRevB.65.165118}, and for group III B
- VI B elemental liquids \cite{PhysRevB.67.104205}.

In recent years, vibration-transit (V-T) theory has been under development to
provide a tractable approximate Hamiltonian for monatomic liquids
\cite{PhysRevE.56.4179, JPhys_13_R739}. In this theory the nuclear motion is
composed of two parts, the many-body vibrational motion in one potential
energy valley, plus transits, which carry the system from valley to valley.
Transits proceed at a high rate throughout the liquid, and are responsible for
diffusion. The vibrational motion is tractable and is subject to \emph{ab
initio} evaluation. Closed-form equations are available for the dominant
structural and vibrational contributions to thermodynamic functions.  The
transit contribution is complicated but small, and is treated by a model.  The
question we ask is simple and direct: If we apply DFT in the adiabatic
approximation to V-T theory, how do the calculated thermodynamic properties
compare with experiment? Our purpose here is to provide an initial answer to
this question.

The study is done for Na and Cu. The properties we calculate are the
equilibrium volume, the isothermal bulk modulus, and the internal energy and
entropy. Comparison of theory and experiment is done for the crystal at $T =
0$ and at the melting temperature $T_{m}$, and for the liquid at $T_{m}$. The
crystal tests are made to establish a fiducial for the theoretical accuracy.
Both Na and Cu have a nearly-free-electron density of states in the vicinity
of the Fermi energy.  Hence the electronic excitation contribution to
thermodynamic properties is very small and may be calculated from free
electron theory. In this way, the ultimate comparison of theory and experiment
is not significantly contaminated by error from electronic excitation. On the
other hand, while Na has a rigid core, with only one valence electron that
deforms as the nucleus moves, both the valence $s$-electron and the filled
$d$-shell deform as the nucleus moves in Cu. This strong difference in the
groundstate adiabatic potential adds dimension to the present study. While
this study is limited to two liquid metals, extensive analysis of experimental
data reveal a common behavior of most elemental liquids \cite{PhysRevE.56.4179},
and Na and Cu are representative of this common behavior.

Before completing the work reported here, two preliminary results were
required.  First, an efficient DFT quench procedure for locating the many-body
potential energy minima was developed \cite{PhysRevE.80.051111,ISI:000280366200004}.  Second, an
accurate predictive model for the transit contribution to thermodynamic
functions was constructed \cite{wallace_statistical_model}. The culmination of
the complete project is reported here.  It must be observed that the present
work is not intended to replace \emph{ab initio} MD calculations; indeed the
two methods are quite complementary, as discussed in the final section.

In Sec.~\ref{sec:II}, we outline the theory and explain various issues
relevant to real-world calculations. In Sec.~\ref{sec:III}, the DFT calculations are
described, including quench procedures and the total energy for crystal and
liquid structures.  (Miscellaneous details related to these sections are
collected in the appendices.) Results are presented and
discussed in Sec.~\ref{sec:IV}. Intermediate theoretical results show the
relative importance of the separate theoretical contributions to internal
energy and entropy. Theory and experiment are tabulated and compared at the
precision of the experimental accuracy. Conclusions are summarized in
Sec.~\ref{sec:V}. The primary conclusion regards the overall accuracy of the
present calculations of thermodynamic properties of liquid Na and Cu. A
secondary conclusion summarizes the verification of V-T theory which is
provided by the present calculations.

\section{Theoretical formulation}
\label{sec:II}
 
The condensed matter potential energy surface is composed of intersecting
many-atom potential energy valleys. Each valley makes a contribution to the
partition function. To approximate the single valley partition function, the
valley potential is harmonically extended to infinity. The partition function
is then simple, at the cost of neglecting anharmonicity and valley-valley
intersections.  We summarize the harmonic single valley statistical mechanical
formulas in Appendix \ref{app:A}.  In this section we show why these
formulas are needed for the present work.

For liquids, the starting point of V-T theory is a hypothesis about the nature of the
many-body potential energy valleys which underlie the nuclear motion. These
valleys are divided into two classes, random and symmetric. In the
thermodynamic limit, the random valleys are supposed to dominate the liquid
statistical mechanics, and are also supposed to be macroscopically uniform.
Macroscopic uniformity means that the statistical mechanical average of any
macroscopic dynamical variable is the same for all random valleys. Therefore
the vibrational contribution to a thermodynamic function can be calculated
from a single random valley harmonically extended to infinity. The
single-valley vibrational motion is supposed to be interspersed with transits,
which carry the system from valley to valley. Hence there are two separate
components of the nuclear motion, vibrations and transits. With a superscript
$l$ to represent the liquid, the total free energy $F^{l} (V, T)$ is written
\begin{equation}
\label{eq:F_l}
F^{l} (V, T) = \Phi^{l}_{0} (V) + F^{l}_{\mathrm{vib}} (V, T) +
F^{l}_{\mathrm{tr}} (V, T) + F^{l}_{\mathrm{el}} (V, T).
\end{equation}
Here $\Phi^{l}_{0} (V)$ is the system potential energy at the random valley
structure, and $F^{l}_{\mathrm{vib}} (V, T)$ and $F^{l}_{\mathrm{tr}} (V, T)$
express respectively the nuclear motion contribution from vibrations and
transits. The final term, $F^{l}_{\mathrm{el}} (V, T)$, represents electronic
excitations; it is added here because Na and Cu, the materials we study, are
metals. However, $F^{l}_{\mathrm{el}} (V, T)$ is very small, and free-electron
theory in the leading Sommerfeld expansion provides sufficient accuracy.

The corresponding internal energy $U^{l} (V, T)$ and entropy $S^{l} (V, T)$
are
\begin{eqnarray}
\label{eq:U_l}
U^{l} (V, T) & = & \Phi^{l}_{0} (V) + U^{l}_{\mathrm{vib}} (V, T) \nonumber \\
 & & + U^{l}_{\mathrm{tr}} (V, T) + U^{l}_{\mathrm{el}} (V, T), \\
\label{eq:S_l}
S^{l} (V, T) & = & S^{l}_{\mathrm{vib}} (V, T) + S^{l}_{\mathrm{tr}} (V, T) +
S^{l}_{\mathrm{el}} (V, T).
\end{eqnarray}
At $T \ge T_{m}$, the melting temperature, the high-$T$ expansions
Eqs.~(\ref{eq:S_vib}) and (\ref{eq:U_vib_expanded}) are valid, hence the
primary potential energy parameters are $\Phi^{l}_{0} (V)$ for the internal
energy and $\theta^{l}_{0} (V)$ for the entropy. $\theta^{l}_{0} (V)$ is the
characteristic temperature related to the log moment of the vibrational
spectrum (see Appendix \ref{app:A}).

A new challenge, unique to the liquid, is to find the structures
corresponding to random valleys so the characteristic functions may be
calculated.  The technique we've developed to do so
\cite{PhysRevE.80.051111,ISI:000280366200004} exploits their numerical dominance; we start
with computer generated \emph{stochastic} configurations, in which the
nuclei are distributed uniformly over the system volume, within a
constraint limiting the closeness of approach of any pair.  As we
demonstrate in \cite{PhysRevE.80.051111,ISI:000280366200004}, quenching from such a
configuration lands the system in a random valley with high
likelihood.

Once the structure is found, the system potential is corrected to the
thermodynamic zero to produce $\Phi^{l}_{0} (V)$ (see Appendix \ref{app:B}).
The dynamical matrix is the mass-weighted curvature tensor evaluated at the
structure. This is calculated by a finite-difference approach in which
each individual nucleus is displaced in all three Cartesian directions
and the forces on all nuclei are computed.  The eigenvalues
are $M \omega_{\lambda}^{2}$ for $\lambda = 1, \dots, 3N$, where $M$
is the nuclear mass and $\omega_{\lambda}$ are the normal mode vibrational
frequencies. Here, to calculate thermodynamic functions, only the eigenvalues
are needed. However, the eigenvectors are also important, as they are needed
to calculate fluctuations and time correlation functions \cite{DCW_PRE08a}.

Although we will perform these calculations in Sec.~\ref{sec:III}
using periodic boundary conditions, that does not imply that the
liquid vibrations correspond to those of a crystal with a large unit
cell.  The liquid vibrational modes are fundamentally different from
those of a crystal.  Since a crystal is periodic in space, periodic
boundary conditions on a crystal unit cell or supercell merely express
the infinite extension of the crystal.  This is the infinite lattice
model \cite{BornHuang}, and entails no error.  A liquid is represented
by a random structure, which has no spatial periodicity, and
calculations for a finite system contain surface errors.  Such errors
are minimized by the application of periodic boundary conditions, but
the resulting spatial periodicity is not physically correct or
meaningful for the liquid.

Finally for the liquid, we must evaluate the transit contributions to the
internal energy and entropy at melt. It was just this requirement, in the
present application of DFT to liquid dynamics theory, that motivated our
development of an improved model for the transit free energy of monatomic
liquids. This model consists of two parts: (a) The available high-$T$
experimental entropy data were analyzed in terms of the entropy formulas,
Eqs.~(\ref{eq:S_l}) and (\ref{eq:S_vib}), revealing a scaled $T$-dependence of
$S^{l}_{\mathrm{tr}} (V, T)$ at fixed volume \cite{PhysRevE.79.051201}, and (b)
a statistical mechanical model for $F^{l}_{\mathrm{tr}} (V, T)$ was calibrated
to this experimental $S^{l}_{\mathrm{tr}} (V, T)$ function, yielding model
equations for all thermodynamic functions which derive from the transit free
energy \cite{wallace_statistical_model}. The model provides universal curves
for $S^{l}_{\mathrm{tr}} / N k_{B}$ and $U^{l}_{\mathrm{tr}} / N k_{B} T$ in
terms of $T / \theta_{\mathrm{tr}} (V)$, where $\theta_{\mathrm{tr}} (V)$ is
the material-specific scaling temperature for the transit entropy. Results for
Na and Cu at melt are listed in Table~\ref{table:I}. Volume dependence of the
transit contribution is neglected.

The temperature $\theta_{\mathrm{tr}} (V)$ plays a role in the transit
contribution similar to that played by $\theta_{0} (V)$ in the
vibrational contribution; it sets a material-specific temperature
scale.  We know how to calculate $\theta_{0} (V)$ from first
principles because the relevant term in the liquid Hamiltonian (the
vibrational part) is well-understood; however, the theory for the
transit contribution to the Hamiltonian is still under development.
Once the transit term is understood, we hope to be able to calculate
$\theta_{\mathrm{tr}} (V)$ from first principles as well.  For now,
parameterization from data will suffice.

\begin{table}
\caption{\label{table:I} Transit contributions to energy and entropy at melt.
\cite{wallace_statistical_model}}
\begin{ruledtabular}
\begin{tabular}{lccccc}
Element
  & $T_{m} / \theta_{\mathrm{tr}}$
  & $U_{\mathrm{tr}} / N k_{B} T_{m}$
  & $S_{\mathrm{tr}} / N k_{B}$ \\
\hline
Na & 0.65 & 0.415 & 0.72 \\
Cu & 1.00 & 0.332 & 0.80 \\
\end{tabular}
\end{ruledtabular}
\end{table}

In the crystal, the system moves in the crystal potential energy valley, and
this motion is well described by lattice dynamics theory \cite{BornHuang}. In
the harmonic approximation, and neglecting valley-valley intersections,
the equations of Appendix \ref{app:A} apply. Therefore, the
total crystal free energy is
\begin{equation}
\label{eq:F_crystal}
F^{c} (V, T) = \Phi^{c}_{0} (V) + F^{c}_{\mathrm{vib}} (V, T) +
F^{c}_{\mathrm{el}} (V, T),
\end{equation}
where $\Phi^{c}_{0} (V)$ is the system potential at the valley minimum, the
crystal structure, and $F^{c}_{\mathrm{vib}} (V, T)$ is the contribution from
lattice vibrations. Again the electronic excitation term $F^{c}_{\mathrm{el}}
(V, T)$ is very small and is given to sufficient accuracy by free electron
theory.

\begin{table}
\caption{\label{table:II} Setup parameters for the VASP calculations.  The
  Monkhorst-Pack $\mathbf{k}$-mesh is recorded as [n, n, n], followed by the
  number of $\mathbf{k}$-points in the irreducible Brillouin zone as
  ($n_{\mathbf{k}}$).  The quantity ``translational invariance'' is the maximum magnitude of
  translational eigenvalues relative to that of the lowest pure vibrational
  mode. ``DM'' is short for ``dynamical matrix.''}
\begin{ruledtabular}
\begin{tabular}{llll}
Quantity & Na & Cu \\
\hline
valence  electrons      & 1         &  11 \\
planewave cutoff [eV]   & 101.7     &  341.6 \\
max core radius [{\AA}] & 2.5       &  2.3 \\
\texttt{EDIFF} [eV]     & $10^{-8}$ & $10^{-8}$ \\
$\mathbf{k}$-mesh for $E^{l} (V)$
  & [3, 3, 3] (14)
  & [5, 5, 5] (63) \\
$\mathbf{k}$-mesh for liquid DM
  & [3, 3, 3] (14)
  & [2, 2, 2] (4) \\
translational invariance (liquid) & $10^{-6}$ & $10^{-6}$ \\
crystal structure          & bcc       & fcc \\
$\mathbf{k}$-mesh for crystal DM
  & [3, 3, 3] (10)
  & [2, 2, 2] (2) \\
translational invariance (crystal) & $10^{-13}$ & $10^{-7}$ \\
\end{tabular}
\end{ruledtabular}
\end{table}

Corresponding to Eq.~(\ref{eq:F_crystal}), the crystal internal energy and
entropy are given by
\begin{eqnarray}
\label{eq:U_c_crystal}
U^{c} (V, T) & = & \Phi^{c}_{0} (V) + U^{c}_{\mathrm{vib}} (V, T) +
U^{c}_{\mathrm{el}} (V, T), \\
\label{eq:S_c_crystal}
S^{c} (V, T) & = & S^{c}_{\mathrm{vib}} (V, T) + S^{c}_{\mathrm{el}} (V, T).
\end{eqnarray}
For Na and Cu, as with most elements, the high-$T$ expansions
Eqs.~(\ref{eq:S_vib}) and (\ref{eq:U_vib_expanded}) are accurate at $T_{m}$
for the crystal as well as the liquid. The fundamental differences in
finite-$N$ errors for crystal and liquid are described in the
following Section.

\section{Electronic structure calculations}
\label{sec:III}

\subsection{Calculations for the liquid}

Our supercell consists of 150 atoms in a cubic box with periodic boundary
conditions. The value $N = 150$ is large enough that finite-$N$ errors are not
serious, and small enough that a sufficient number of total energy
calculations (a few thousand) can be done for each element.

The DFT calculations are done with the VASP code \cite{VASP}, using the
projector augmented wave (PAW) method \cite{PhysRevB.50.17953} in the
generalized gradient approximation (GGA) \cite{PhysRevB.59.1758}. The
$\mathbf{k}$-point mesh was automatically generated using the method of
\citet{PhysRevB.13.5188}. The setup parameters are listed in
Table~\ref{table:II}.  It is the large size of the real-space supercell which
allows us to use few $\mathbf{k}$-points in comparison to the large number
(several thousands) needed for crystal metal calculations with small unit
cells.

To locate structures, we followed the procedure outlined in Sec.~\ref{sec:II}
\cite{PhysRevE.80.051111,ISI:000280366200004}. A quench is considered 
converged when the energy decrease in one iteration is less than 10
times the current \texttt{EDIFF} setting, where \texttt{EDIFF} defines
the energy convergence criterion used in the SCF procedure (see 
Table~\ref{table:II}). Initially, quenches were done from a separate 
stochastic configuration at each volume. To save computer time we
quenched to one structure, then scaled this structure to slightly larger and
smaller volumes, and quenched these configurations to new structures. This was
repeated until the desired range of volumes was covered.  In quenching with
DFT, the following procedure saves computer time: quench to convergence at
Monkhorst-Pack $\mathbf{k}$-mesh [1, 1, 1], then take Monkhorst-Pack
$\mathbf{k}$-mesh [2, 2, 2] and quench to convergence again, and so on. This
is faster, not because it requires fewer quench steps, but because most of the
steps are at a smaller number of $\mathbf{k}$-points.

\begin{figure}
\caption{\label{fig:I}For Na at $N = 150$: DFT results for the structural
energy $E$ versus volume $V$ for the liquid ($E^{l}$), and for the crystal
($E^{c}$).}
\includegraphics*[angle=0, width=\linewidth]{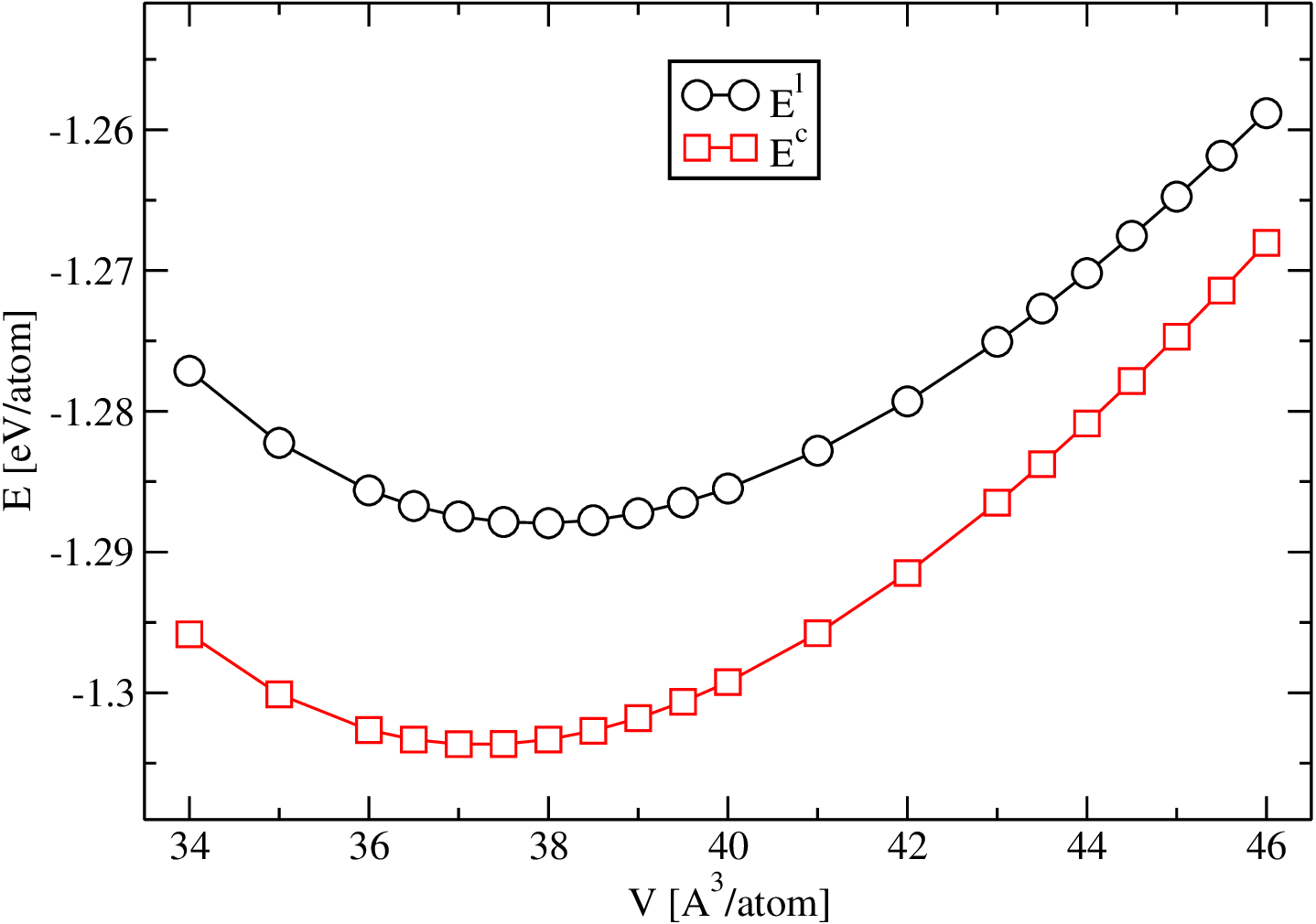}
\end{figure}

\begin{figure}
\caption{\label{fig:II}For Cu at $N = 150$: DFT results for the structural
energy $E$ versus volume $V$ for the liquid ($E^{l}$), and for the crystal
($E^{c}$).}
\includegraphics*[angle=0, width=\linewidth]{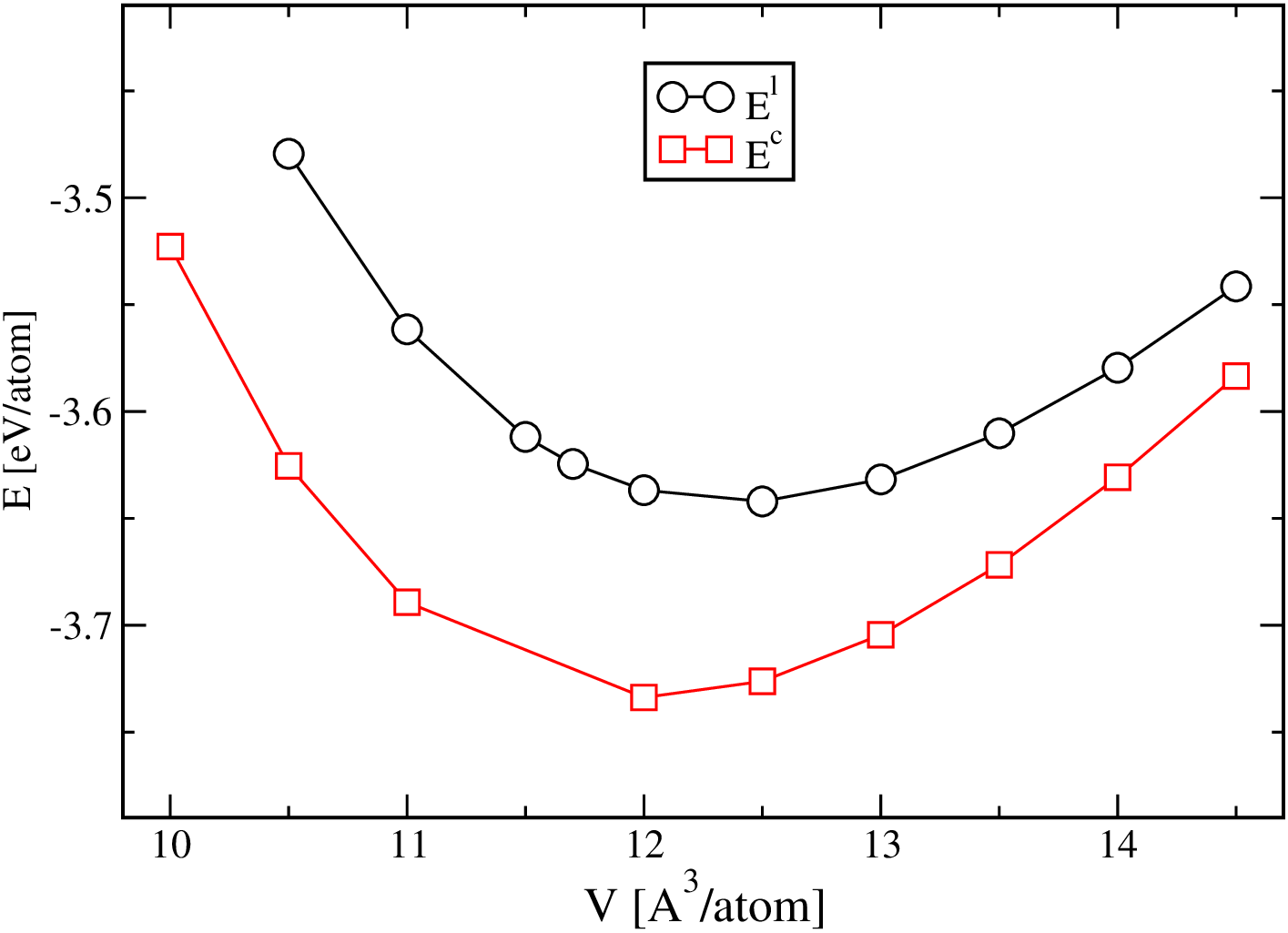}
\end{figure}

The DFT energy of a random structure is denoted $E^{l} (V)$. The results
for Na and Cu are shown in Figs.~\ref{fig:I} and \ref{fig:II} respectively.
The crystal structure energies $E^{c} (V)$ are also shown in the figures. From
the potential energy hypothesis mentioned in Sec.~\ref{sec:II}, for a given
element at a given $V$, the random structural energies should occupy a
distribution whose width is small compared to $k_{B} T_{m}$, and whose mean
lies above $E^{c}$ by around $k_{B} T_{m}$. This characteristic is well
verified for Na at the volume $V^{l}_{m}$ of the liquid at melt
\cite{PhysRevE.80.051111, PRE_59_2955, PRE_76_041203}. Here, since each $E^{l}
(V)$ is a representative of the random distribution at that $V$, the curves of
$E^{l} (V)$ and $E^{c} (V)$ in Fig.~\ref{fig:I} confirm the characteristic
random structure distribution over a range of volumes for Na. The same
confirmation is provided for Cu in Fig.~\ref{fig:II}.  These energies
are then normalized as described in Appendix \ref{app:B} to provide
the structural potential for the liquid, $\Phi^{l}_{0} (V)$.

In the many quenches done in the present study, very few symmetric structures
have appeared. These structures invariably have DFT energies lying noticeably
below the curve of $E^{l} (V)$, and above $E^{c} (V)$. This also accords with
the potential energy hypothesis of V-T theory, Sec.~\ref{sec:II}, and accords
with previous findings \cite{PhysRevE.80.051111, PRE_59_2955, PRE_76_041203}
for Na at $V^{l}_{m}$.

For a given liquid at fixed density, the macroscopic random structure is
better approximated with an ever larger supercell. Hence at finite $N$ all
calculated potential energy parameters will have an error due to finite
resolution of the structure. Moreover the vibrational characteristic temperatures are
subject to a second finite-$N$ error, due to the incomplete resolution of the
frequency distribution. This second error is the dominant error in our liquid
characteristic temperature calculations.

Among the normal modes are three, $\lambda = 1, 2, 3$, representing uniform
translation of the system. Their eigenvalues are in principle zero due
to translational invariance (also called the acoustic sum rule in crystal
theory). In practice the translational eigenvalues are zero only to numerical
accuracy, and are of either sign. The remaining $3N-3$ modes are pure
vibrational and have positive eigenvalues, by the definition of a structure as
a local minimum. To check translational invariance, the ratio of eigenvalues
$\left| \omega^{2}_{\lambda} \right| / \omega^{2}_{4}$ is calculated for
$\lambda = 1, 2, 3$, where mode $\lambda = 4$ is the lowest lying pure
vibrational mode. The maximum value of this ratio in our final calculations
for each element is listed under the designation ``translational invariance''
in Table~\ref{table:II}. The requirement is clearly satisfied to high
numerical accuracy.

\begin{figure}
\caption{\label{fig:III}For liquid Na at $N = 150$: DFT results for the
vibrational characteristic temperatures $\theta_{n}$ versus volume $V$,
for $n = 0, 1, 2$ (open symbols). Also the same $\theta_{n}$ at $N = 500$ and
$V = V^{l}_{m}$, from Na interatomic potentials \cite{PRE_76_041203} (solid
symbols).}
\includegraphics*[angle=0, width=\linewidth]{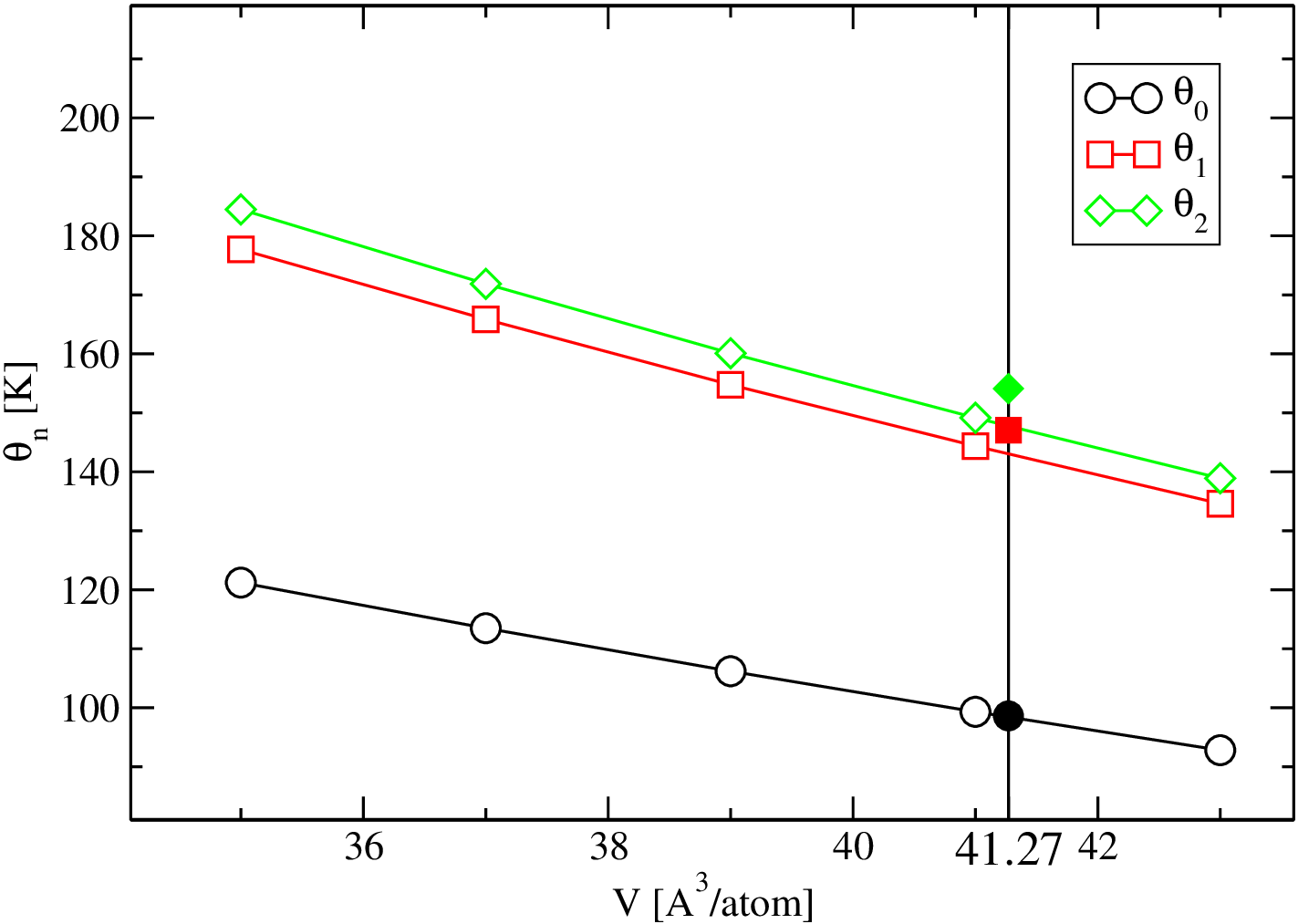}
\end{figure}

Here we are interested in moments of the frequency distribution, which are related to
characteristic temperatures $\theta_{n}$ by Eqs.~(\ref{eq:moment_0}) -
(\ref{eq:moment_2}), where the translational eigenvalues are excluded
from the average.  Our calculations of these $\theta_{n}$ for Na are shown
in Fig.~\ref{fig:III}.  Also shown are values from a well-tested interatomic
potential for Na, for a 500 atom system at the volume of the liquid at melt
\cite{PRE_76_041203}. Agreement is excellent for $\theta_{0}$, and is good for
$\theta_{1}$ and $\theta_{2}$. It is common for theory and experiment both to
give more reliable values for $\theta_{0}$ than for other $\theta_{n}$. The
reason is that $\theta_{0}$, being the log moment, is uniformly sensitive to
all $\omega_{\lambda}$ in the spectrum, while other $\theta_{n}$ are more
sensitive to higher (or lower) frequencies, hence depend on a smaller portion
of the spectrum.

\subsection{Calculations for the crystal}

For the structural energy calculations, we use one primitive unit cell with
periodic boundary conditions, and increase the $\mathbf{k}$-mesh to
convergence.  This represents the infinite lattice model \cite{BornHuang}, and
there are no finite-$N$ errors. The situation contrasts with the liquid
calculation, where the structure itself contains a finite-$N$ error.

For the crystal, our calculated phonon moments can be tested against results
from inelastic neutron scattering \cite{SchoberDederichs}. This will
ultimately allow us to estimate the accuracy of the calculated liquid moments.
To this end, we calculate the crystal dynamical matrix for a large supercell
with periodic boundary conditions, just as for the liquid. The procedure
yields phonons with wavevectors commensurate with the supercell. The crystal
structure is precise, but the phonon moments have finite-$N$ error due to the
limited resolution of the frequency distribution. Since this is also the major
error in the liquid, we expect the total error in vibrational characteristic
temperatures to be about the same for crystal and liquid at a given $N$.

To calculate the dynamical matrix, a rhombohedral supercell of $5 \times 5
\times 5$ primitive unit cells was constructed for bcc and fcc lattices
\cite{AshcroftMermin}.  This gives $N = 125$, as close as
possible to the liquid value of $N$. A fixed $\mathbf{k}$-mesh was chosen, the
same as that used for the liquid; the same $\mathbf{k}$-mesh produces a
smaller number of $\mathbf{k}$-points in the irreducible Brillouin zone for
the crystal (see Table~\ref{table:II}).

\section{Results}
\label{sec:IV}

\subsection{Intermediate theoretical results}

\begin{table}
\caption{\label{table:III} Intermediate Theoretical Results}
\begin{ruledtabular}
\begin{tabular}{llll}
Quantity & Na & Cu \\
\hline
\underline{Crystal at $T = 0$} & & \\
$V^{c}_{\mathrm{ref}}$ [{\AA}${}^{3}$/atom]    & 37.24   &  12.02 \\
$D$ [eV/atom]                                  & 1.28763 &  3.7047 \\
$\frac{9}{8} N k_{B} \theta_{1}$ [meV/atom]    & 16.05   &  29.3 \\
$\frac{9}{8} k_{B} \theta_{1} / 3 k_{B} T_{m}$ & 0.167   &  0.087 \\
\underline{Crystal at $T_{m}$} & & \\
$V^{c}_{m}$ [{\AA}${}^{3}$/atom]      &  39.79 &   13.05 \\
$\Phi^{c}_{0}$ [meV/atom]             & -12.22 &    3.1  \\
$U^{c}_{\mathrm{vib}}$ [meV/atom]     &  96.72 &  351.7  \\
$U^{c}_{\mathrm{el}}$ [meV/atom]      &   0.82 &   5.3   \\
$S^{c}_{\mathrm{vib}}$ [$k_{B}$/atom] &   6.79 &   9.03  \\
$S^{c}_{\mathrm{el}}$ [$k_{B}$/atom]  &   0.05 &   0.09  \\
\underline{Liquid at $T_{m}$} & & & \\
$V^{l}_{m}$ [{\AA}${}^{3}$/atom]      &  40.93 &   13.54 \\
$\Phi^{l}_{0}$ [meV/atom]             &   4.62 &   96.5  \\
$U^{l}_{\mathrm{vib}}$ [meV/atom]     &  96.65 &  351.7  \\
$U^{l}_{\mathrm{tr}}$ [meV/atom]      &  13.27 &   38.9  \\
$U^{l}_{\mathrm{el}}$ [meV/atom]      &   0.82 &   5.2   \\
$S^{l}_{\mathrm{vib}}$ [$k_{B}$/atom] &   6.96 &   9.39  \\
$S^{l}_{\mathrm{tr}}$ [$k_{B}$/atom]  &   0.72 &   0.80  \\
$S^{l}_{\mathrm{el}}$ [$k_{B}$/atom]  &   0.05 &   0.09  \\
\end{tabular}
\end{ruledtabular}
\end{table}

For the crystal at $T = 0$ and $T_{m}$, and for the liquid at $T_{m}$, we fit
the four-parameter Vinet-Rose function \cite{0022-3719-19-20-001} to our
calculated $F (V, T)$ versus $V$. From this we find the volume at $P = 0$, and
the isothermal bulk modulus $B$ at that volume. Intermediate theoretical
results calculated from the formulas of Sec.~\ref{sec:II} are listed in
Table~\ref{table:III}. These show the relative importance of various
theoretical contributions to internal energy and entropy. The discussion here
is for Na and Cu collectively, and is qualitatively applicable to monatomic
crystals and liquids in general.

At $T = 0$, the free energy is given by Eqs.~(\ref{eq:crystal_free_energy}) -
(\ref{eq:energy_constant}). For accurate theoretical work, the zero-point
energy cannot be neglected. For the light elements the zero-point energy
measurably affects the volume at $P = 0$ (see also Ref.~\cite{Wallace_SPCL},
Table~16.3). For all elements, the zero-point energy is important in the
internal energies of crystal and liquid states.  This is shown by the ratio of
the zero-point energy to the classical vibrational energy at melt,
$\frac{9}{8} k_{B} \theta^{c}_{1} / 3 k_{B} T_{m}$, listed in
Table~\ref{table:III}.  If the zero-point energy is omitted from theory, this
ratio is the relative error made in the (dominant) internal energy
contribution $U_{\mathrm{vib}} (V, T)$.

For the crystal at melt, the internal energy and entropy are given by
Eqs.~(\ref{eq:U_c_crystal}) and (\ref{eq:S_c_crystal}). From
Table~\ref{table:III}, the dominant energy contribution is
$U^{c}_{\mathrm{vib}}$. The small contribution from $\Phi^{c}_{0}$ is a
combination of the volume-dependent part of $E^{c} (V)$, and the zero-point
energy, according to Eqs.~(\ref{eq:crystal_structure_energy}) and
(\ref{eq:energy_constant}). The contribution $E^{c} (V) - E^{c}
(V^{c}_{\mathrm{ref}})$ can be read from Figs.~\ref{fig:I} and \ref{fig:II}.
The dominant entropy contribution is again vibrational. At $T \gtrsim T_{m}$,
$S^{c}_{\mathrm{vib}}$ depends almost entirely on $T / \theta^{c}_{0} (V)$,
from Eq.~(\ref{eq:S_vib}). In Table~\ref{table:III}, the electronic
contributions are quite small, being $\lesssim$ 2\% for the crystal at melt.
These contributions are much larger, say up to 10\%, for metals with unfilled
$d$-bands \cite{PhysRevB.46.5221}.

\begin{table*}
\caption{\label{table:IV} Comparison of theory and experiment for the crystal
  at low $T$. $V^{c}_{\mathrm{ref}}$ and $B^{c}$ are at $T = 0$ and $P = 0$.
  $\theta^{c}_{n}$ are at $V^{c}_{\mathrm{meas}}$, the volume of experimental
  measurement.}
\begin{ruledtabular}
\begin{tabular}{llllllllll}
\multirow{2}{*}{Quantity}
  & \multicolumn{3}{c}{Na (bcc)}
  & \multicolumn{3}{c}{Cu (fcc)} \\
  & Theory
  & Expt.
  & $\Delta$
  & Theory
  & Expt.
  & $\Delta$ \\
\hline
$V^{c}_{\mathrm{ref}}$ [{\AA}${}^{3}$/atom]
  & 37.24
  & 37.68\footnote{Tables~15.1 and 19.1 of Ref.~\cite{Wallace_SPCL}}\setcounter{footnote1}{\value{mpfootnote}}
  & -0.012
  & 12.02
  & 11.70\footnotemark[\value{footnote1}]
  & 0.027 \\
$B^{c}$ [GPa]
  & 7.76
  & 7.3-7.6\footnote{See Refs.~\cite{PhysRevB.28.5395, PhysRevB.31.668, Diederich1966637, PhysRev.178.902}}
  & 0.042
  & 138.7
  & 142.0\footnote{See Ref.~\cite{SimmonsWang}}\setcounter{footnote5}{\value{mpfootnote}}
  & -0.023 \\
\hline
$V^{c}_{\mathrm{meas}}$ [{\AA}${}^{3}$/atom]
  & 37.98
  & 37.98\footnotemark[\value{footnote1}]
  & ---
  & 11.70
  & 11.70\footnotemark[\value{footnote1}]
  & --- \\
$\theta^{c}_{0}$ [K]
  & 111.2
  & 113.3\footnote{See Ref.~\cite{SchoberDederichs} and Tables~15.1 and 19.1 of Ref.~\cite{Wallace_SPCL}}\setcounter{footnote61}{\value{mpfootnote}}
  & -0.019
  & 236.5
  & 225.3\footnotemark[\value{footnote61}]
  & 0.050 \\
$\theta^{c}_{1}$ [K]
  & 161.7
  & 163\footnotemark[\value{footnote61}]
  & -0.008
  & 330.2
  & 315\footnotemark[\value{footnote61}]
  & 0.048 \\
$\theta^{c}_{2}$ [K]
  & 165.8
  & 166\footnotemark[\value{footnote61}]
  & -0.001
  & 332.5
  & 317\footnotemark[\value{footnote61}]
  & 0.049 \\
\end{tabular}
\end{ruledtabular}
\end{table*}

The liquid thermodynamic functions contain terms analogous to those in the
crystal, plus an added contribution from transits,
Eqs.~(\ref{eq:F_l}) - (\ref{eq:S_l}) and (\ref{eq:energy_constant_liquid}). For
the liquid at melt, the character of contributions from the structural
potential $\Phi^{l}_{0}$, from vibrations, and from electronic excitations is
qualitatively the same as described above for the crystal at melt. Again the
contribution from $E^{l} (V) - E^{c} (V^{c}_{\mathrm{ref}})$ can be read from
Figs.~\ref{fig:I} and \ref{fig:II}.  However, the transit contribution in
Table~\ref{table:III} is approximately 10\% for the liquid at melt, and is
therefore important for an accurate theory. This is the only entry in
Table~\ref{table:III} not obtained from electronic structure calculations. But
this term also will be amenable to DFT calculation, as soon as a model for the
transit Hamiltonian is developed.

\subsection{Comparison of theory and experiment}

Comparison of theory and experiment for the crystal at $T = 0$ and $P = 0$ is
listed in Table~\ref{table:IV}. Differences are expressed in the quantity
$\Delta$, defined in general by
\begin{equation}
\Delta = \frac{\mbox{theory} - \mbox{expt}}{\mbox{expt}}.
\end{equation}
For the crystal volume and bulk modulus, the agreement is excellent, at the
customary level for \emph{ab initio} crystal calculations (see e.g.
Ref.~\cite{Wallace_SPCL}, Tables~16.2 and 16.3).

In Table~\ref{table:IV}, comparison of $\theta^{c}_{n} (V)$ for $n = 0, 1, 2$
is at the experimental volume $V^{c}_{\mathrm{meas}}$ (see
Ref.~\cite{Wallace_SPCL}, Table~15.1). The experimental error in
$\theta^{c}_{n} (V^{c}_{\mathrm{meas}})$ is estimated to be 0.1 - 0.5\% (p.~151 
of Ref.~\cite{Wallace_SPCL}). In our experience, lattice dynamics theory
in \emph{ab initio} evaluation can account for the experimental
$\theta^{c}_{n}$ to an accuracy around 1\% at best. To achieve such accuracy
is not our goal here. Accuracy of the \emph{ab initio} $\theta^{c}_{n}$ in
Table~\ref{table:IV} is quite respectable, with $\Delta$ in the range $-2$\%
to $+5$\%. Error at this level is a minor effect in the comparison of theory
and experiment for the crystal at melt. Moreover, we attribute the theoretical
error mainly to small system size ($N = 125$), since the corresponding small
number of vibrational modes can only poorly represent the actual crystal
frequency distribution. This problem is easily remedied by increasing $N$.

Comparison of theory and experiment for the crystal at melt, and for the
liquid at melt, is listed in Table~\ref{table:V}. Notice the nuclear motion
causes thermal expansion, as seen in the volume at melt. Specifically,
$V^{c}_{m}$ is larger than $V$ at the minimum of $E^{c} (V)$, and $V^{l}_{m}$
is larger than $V$ at the minimum of $E^{l} (V)$, Figs.~\ref{fig:I} and
\ref{fig:II}. The volumes are in excellent agreement with experiment. Also, at
this point we can see that the volume errors are systematic: for all three
states, the crystal at $T = 0$ and at $T_{m}$, and the liquid at $T_{m}$, the
volume error is $-0.01$ for Na, and $+0.03$ for Cu. The bulk modulus, being
essentially the curvature of $\Phi_{0} (V)$, for crystal or for liquid, has
larger error than the volume itself. Moreover, the experimental and
theoretical determinations of $B$ in Table~\ref{table:V} are \emph{both}
subject to significant errors. The agreement of theory and experiment for $B$
is as good as we can expect at $T_{m}$.

It remains to discuss the comparison of theory and experiment for the internal
energy and entropy at melt, Table~\ref{table:V}. The errors here are
sufficiently small that individual $\Delta$ values cannot be interpreted. The
mean and standard deviation of $\Delta$ for energy and entropy for crystal and
liquid states is $\Delta = -0.005 \pm 0.019$. Hence the errors are essentially
pure scatter. Contributions to the scatter arise from two major sources,
experimental error at the level of 0.005 - 0.010, and computational error due
to small system size at the level of 0.01 - 0.02. Additional smaller errors
result from the slightly inaccurate theoretical volume, from error in the
free-electron model for electronic excitation, and from neglect of electronic
excitation-nuclear motion interaction \cite{PhysRevB.72.155120, bock:075114}.
It follows that the results in Table~\ref{table:V} for internal energy and
entropy at melt are consistent with known errors.

There is one more systematic property of the comparison in Table~\ref{table:V}
that holds separately for each metal. For the bulk modulus, $\Delta$ is large
and positive and is roughly the same for crystal and liquid, while for each
remaining property, $\Delta$ is small and approximately the same magnitude for
crystal and liquid. The implication is that the liquid theory we study has the
same level of accuracy as does lattice dynamics theory for crystals.

\begin{table*}
\caption{\label{table:V} Comparison of theory and experiment for crystal and
  liquid at $T_{m}$}
\begin{ruledtabular}
\begin{tabular}{llllllllll}
\multirow{2}{*}{Quantity}
  & \multicolumn{3}{c}{Na (bcc)}
  & \multicolumn{3}{c}{Cu (fcc)} \\
  & Theory
  & Expt.
  & $\Delta$
  & Theory
  & Expt.
  & $\Delta$ \\
\hline
\underline{Crystal Data} & & & & & & & \\
$V^{c}$ [{\AA}${}^{3}$/atom]
  & 39.79
  & 40.27\footnote{Tables~19.1 and 21.1 of Ref.~\cite{Wallace_SPCL}}\setcounter{footnote1}{\value{mpfootnote}}
  & -0.012
  &  13.05
  &  12.62\footnotemark[\value{footnote1}]
  &  0.034 \\
$B^{c}$ [GPa]
  &  6.47
  &  5.8\footnote{See Ref.~\cite{Fritsch19731961}}
  &  0.116
  &  97
  &  90\footnote{See Ref.~\cite{SimmonsWang}}\setcounter{footnote5}{\value{mpfootnote}}
  &  0.078 \\
$U^{c}$ [meV/atom]
  & 85.32
  & 89.1\footnote{See Ref.~\cite{Hultgren}}\setcounter{footnote8}{\value{mpfootnote}}
  & -0.042
  & 353.1
  & 358.9\footnotemark[\value{footnote8}]
  & -0.016 \\
$S^{c}$ [$k_{B}$/atom]
  &  6.84
  &  6.93\footnotemark[\value{footnote8}]
  & -0.013
  &   9.12
  &   8.93\footnotemark[\value{footnote8}]
  &  0.021 \\
\hline
\underline{Liquid Data}     & & & & & & \\
$V^{l}$ [{\AA}${}^{3}$/atom]
  &  40.93
  &  41.27\footnotemark[\value{footnote1}]
  & -0.008
  &  13.54
  &  13.19\footnotemark[\value{footnote1}]
  & 0.027 \\
$B^{l}$ [GPa]
  &   5.93
  &   5.3\footnote{See Refs.~\cite{HighTemp_4_352, PhysRevB.32.7937}}
  &  0.119
  &  85.7
  &  73.6\footnote{See Ref.~\cite{PhysPropLiqMet}}
  & 0.164 \\
$U^{l}$ [meV/atom]
  & 115.36
  & 115.9\footnotemark[\value{footnote8}]
  & -0.005
  & 492.3
  & 492.7\footnotemark[\value{footnote8}]
  & -0.001 \\
$S^{l}$ [$k_{B}$/atom]
  &   7.73
  &   7.78\footnotemark[\value{footnote8}]
  & -0.006
  &  10.28
  &  10.09\footnotemark[\value{footnote8}]
  & 0.019 \\
\end{tabular}
\end{ruledtabular}
\end{table*}

\section{Conclusions}
\label{sec:V}

\subsection{The present application of DFT to liquid dynamics}

Our primary conclusion from the present study is: Using a standard
implementation of DFT, as provided by the VASP code, it is possible to make
\emph{ab initio} calculations of thermodynamic properties of monatomic liquids.
The liquid calculations achieve the same level of accuracy as does \emph{ab
initio} lattice dynamics for crystals, the most accurate crystal theory
available. This conclusion is validated for Na and Cu, based on the
comparisons of theory with experiment listed in Table~\ref{table:V}. The
following discussion adds detail to the primary conclusion.

The comparisons for the liquid in Table~\ref{table:V} are at zero pressure and
temperature $T_{m}$. However, we fully expect the agreement of theory and
experiment to hold for a wide range of pressures and temperatures, or
equivalently, a wide range of volumes and temperatures. The volume dependence
is contained in functions calculated by DFT, primarily $\Phi^{l}_{0} (V)$ and
$\theta^{l}_{0} (V)$. We can expect these functions to be as accurate for the
compressed liquid as for the liquid at zero pressure. Moreover, at the fixed
volume $V^{l}_{m}$, the temperature dependence of the experimental
thermodynamic data is accurately accounted for by the equations of
Sec.~\ref{sec:II} and Appendix \ref{app:A}. This is because those
equations, and the experimental data for entropy at high temperatures,
were used to calibrate the statistical mechanical model for the
transit free energy \cite{PhysRevE.79.051201}. Hence the level of
agreement between theory and experiment found in Table~\ref{table:V} 
should persist to higher pressures and temperatures.

A potentially useful comparison can be made between the present technique and
\emph{ab initio} MD. At a given $N$, \emph{ab initio} MD requires far greater
computer resources to calculate a similar set of thermodynamic data. Or, with
a fixed computer resource, the present technique can study a larger system,
and hence obtain greater accuracy by reducing finite-$N$ errors. On the other
hand, \emph{ab initio} MD data contain both vibrational and transit
contributions. Both techniques together can separate the transit and
vibrational contributions to statistical mechanical functions. The combined
techniques have the potential to reveal the physical nature of transit motion.

What is achieved by making DFT calculations for the crystal at $T = 0$? First,
before we can compare theory and experiment for any condensed matter state, it
is necessary to adjust the DFT energy calculations to have the thermodynamic
zero of energy. This is accomplished by means of
Eqs.~(\ref{eq:crystal_structure_energy}) and (\ref{eq:energy_constant}).  A
second useful result is the confirmation that DFT calculations are accurate
for $\Phi^{c}_{0} (V)$, and for the characteristic temperatures
$\theta^{c}_{0}$, $\theta^{c}_{1}$, and $\theta^{c}_{2}$. This is shown by the
comparisons of Table~\ref{table:IV}. This confirmation lends confidence to
similar calculations for the liquid.

What is achieved by comparing theory and experiment for the crystal at
$T_{m}$?  This comparison is rarely performed in crystal physics research, and
is of interest in itself. The comparisons of Table~\ref{table:V}
confirm that the DFT calculations are accurate for $\Phi^{c}_{0} (V)$ and
$\theta^{c}_{0} (V)$, \emph{and} their volume derivatives. This confirmation
shows the level of accuracy which \emph{ab initio} lattice dynamics can achieve
for the crystal at melt, and it also lends confidence to the similar
calculations for the liquid.

\subsection{Verification of V-T theory}

The primary conclusion is based on the comparison of theoretical and
experimental data, and uses no condition on the validity of theory. But the
well-established reliability of DFT calculations supports a secondary
conclusion regarding the theory itself. The theory consists of two parts, the
V-T equations and DFT calculations. We can reasonably assume that DFT is as
accurate for the liquid as for the crystal. Then, at this level of accuracy,
the comparisons of theory and experiment for the liquid, in
Table~\ref{table:V}, confirm the equations of V-T theory as described in
Sec.~\ref{sec:II}. Moreover, confirmation of the equations provides
confirmation of the nuclear motion described by the equations. The following
discussion adds detail to this secondary conclusion.

In the formula for internal energy, Eq.~(\ref{eq:U_l}), the major contribution
is $\Phi^{l}_{0} (V) + U^{l}_{\mathrm{vib}} (V, T)$. This is a purely
theoretical function, and describes normal-mode vibrational motion of the
nuclei. Also in Eq.~(\ref{eq:U_l}), no significant error is contributed by the
small terms $U^{l}_{\mathrm{tr}} (V, T)$ and $U^{l}_{\mathrm{el}} (V, T)$.
Agreement of theory and experiment for the internal energy for one liquid
demonstrates consistency of the vibrational motion with experiment. The same
argument applies to the liquid entropy, Eq.~(\ref{eq:S_l}), where the same
nuclear motion is described by the purely theoretical function
$S^{l}_{\mathrm{vib}} (V, T)$.  Agreement of theory and experiment for the
entropy for one liquid demonstrates consistency of the vibrational motion with
experiment. Further, the quantities calculated by DFT, $\Phi^{l}_{0} (V)$ for
the internal energy and $\theta^{l}_{0} (V)$ for the entropy, are not
theoretically related, so the confirmations obtained from internal energy and
entropy are independent.

This situation is analogous when one compares theory with experiment for the
liquid volume and bulk modulus. The theoretical functions tested are volume
derivatives of $\Phi^{l}_{0} (V)$ and $\theta^{l}_{0} (V)$, and the comparison
with experiment for $V^{l}_{m}$ and $B (V^{l}_{m}, T_{m})$ are independent
tests.

Altogether in Table~\ref{table:V}, four independent consistency tests of the
nuclear motion are provided for each of two elemental liquids. The agreement
of theory and experiment within small errors for all these tests provides the
following two-part verification of V-T theory (for Na and Cu): (a) Many-body
harmonic vibrational motion described by the parameters $\Phi^{l}_{0} (V)$ and
$\theta^{l}_{0} (V)$ is the dominant contribution to the thermodynamic
functions, and (b) While the experimental liquid moves rapidly among all the
valleys in the potential energy surface, a single random valley at each volume
serves to calculate the vibrational motion and its contribution to
thermodynamics.

\appendix

\section{Statistical mechanics for a single potential energy valley}
\label{app:A}

The system has $N$ atoms in a volume $V$ at temperature $T$, and is confined
to a single harmonic potential valley. The system potential energy at the
valley minimum, the structure, is $\Phi_{0} (V)$. The normal mode vibrational
frequencies are $\omega_{\lambda} (V)$ for $\lambda = 4, \ldots, 3N$, where
the three translational modes are omitted from statistical mechanics. The
Helmholtz free energy is $F$, given by
\begin{eqnarray}
\label{eq:free_energy}
F (V, T)                & = & \Phi_{0} (V) + F_{\mathrm{vib}} (V, T), \\
\label{eq:F_vib}
F_{\mathrm{vib}} (V, T) & = & \sum_{\lambda} \left[ \frac{1}{2} \hbar \omega_{\lambda}
  - k_{B} T \ln \left( n_{\lambda} + 1 \right) \right],
\end{eqnarray}
where $n_{\lambda}$ is the Bose-Einstein distribution function. In low- and
high-$T$ regimes, $F_{\mathrm{vib}}$ depends on only a few characteristic
temperatures $\theta_{n}$, which are related to moments of the frequency
distribution. Here the important $\theta_{n}$ are given by
\begin{eqnarray}
\label{eq:moment_0}
\ln (k_{B} \theta_{0}) & = & \left< \ln (\hbar \omega) \right>, \\
\label{eq:moment_1}
k_{B} \theta_{1}       & = & \frac{4}{3} \left< \hbar \omega \right>, \\
\label{eq:moment_2}
k_{B} \theta_{2}       & = & \sqrt{ \frac{5}{3} \left< \left( \hbar \omega \right)^{2} \right>},
\end{eqnarray}
where the average $\left< \cdots \right>$ is over the vibrational frequencies.

Functions useful for comparing with experimental data are the internal energy
$U (V, T)$, and the entropy $S (V, T)$. These are obtained from the free
energy, and Eq.~(\ref{eq:free_energy}) yields
\begin{eqnarray}
S (V, T) & = & S_{\mathrm{vib}} (V, T), \\
\label{eq:internal_energy}
U (V, T) & = & \Phi_{0} (V) + U_{\mathrm{vib}} (V, T).
\end{eqnarray}
We shall be interested in the vibrational contributions only at $T = 0$ and
$T_{m}$. At $T = 0$,
\begin{eqnarray}
S_{\mathrm{vib}} (V, T = 0) & = & 0, \\
\label{eq:U_vib}
U_{\mathrm{vib}} (V, T = 0) & = & \frac{9}{8} N k_{B} \theta_{1} (V).
\end{eqnarray}
The right side of Eq.~(\ref{eq:U_vib}) is just the harmonic zero-point energy.
This equation will be needed below to evaluate the thermodynamic zero of
energy. For most monatomic crystals and liquids, the nuclear motion is nearly
classical at $T \gtrsim T_{m}$, and a high-$T$ expansion of $n_{\lambda}$ in
Eq.~(\ref{eq:F_vib}) is valid. This expansion gives
\begin{eqnarray}
\label{eq:S_vib}
S_{\mathrm{vib}} (V, T) & = & 3 N k_{B} \left[ \ln \left( \frac{T}{\theta_{0} (V)}
  \right) + 1 \right. \nonumber \\
 & & \left. + \frac{1}{40} \left( \frac{\theta_{2} (V)}{T} \right)^{2} +
  \cdots \right], \\
\label{eq:U_vib_expanded}
U_{\mathrm{vib}} (V, T) & = & 3 N k_{B} T \left[ 1 + \frac{1}{20} \left(
  \frac{\theta_{2} (V)}{T} \right)^{2} + \cdots \right]. \nonumber \\
\end{eqnarray}
The series starting with $\left( \theta_{2} / T \right)^{2}$ expresses the
quantum corrections, and only this leading term is required in the present
study.

\section{Normalizing DFT energies for thermodynamics}
\label{app:B}

To evaluate the functions in Appendix \ref{app:A}, we require the
potential energy at the valley minimum $\Phi_{0} (V)$ and the moments
of the vibrational spectrum $\theta_{n}$ for $n = 0, 1, 2$.
Extraction of the moments from DFT is described in Sec.~\ref{sec:II};
$\Phi_{0} (V)$ is determined as follows.

All system energies are to be measured from the thermodynamic zero of energy,
which is the energy of the crystal at $T = 0$ and $P = 0$. Here $P$ is
pressure, given by $P = -\left( \partial F / \partial V \right)_{T}$. At $T =
0$, the crystal free energy is
\begin{equation}
\label{eq:crystal_free_energy}
F^{c} (V) = U^{c} (V) = \Phi^{c}_{0} (V) + \frac{9}{8} N k_{B}
\theta^{c}_{1} (V).
\end{equation}
DFT calculations provide total energies $E (V)$ measured with respect to
isolated atoms. To correct the $E (V)$ to the thermodynamic energy zero, we
denote the DFT crystal structure energy by $E^{c} (V)$, and define the
constant $D$ by
\begin{equation}
\label{eq:crystal_structure_energy}
\Phi^{c}_{0} (V) = E^{c} (V) + D.
\end{equation}
If $V^{c}_{\mathrm{ref}}$ is the crystal volume at $T = 0$ and $P = 0$, then
by definition $U^{c} (V^{c}_{\mathrm{ref}}, T = 0) = 0$.
Equations (\ref{eq:crystal_free_energy}) and (\ref{eq:crystal_structure_energy})
can be solved for $D$ to find
\begin{equation}
\label{eq:energy_constant}
D = -\left[ E^{c} (V^{c}_{\mathrm{ref}}) + \frac{9}{8} N k_{B} \theta^{c}_{1}
(V^{c}_{\mathrm{ref}}) \right].
\end{equation}
Equations (\ref{eq:crystal_structure_energy}) and (\ref{eq:energy_constant}) define
$\Phi^{c}_{0} (V)$ in terms of quantities calculated by DFT. To observe the
thermodynamic energy zero for a given atomic system, the same $D$ is added to
every energy calculated in DFT, for every condensed matter phase, at all $V$.
With this, only DFT energy differences enter the quantities we need to
calculate.

As for the crystal valley, potential energy properties of random valleys
depend on the system volume. Hence to include volume dependence, a
representative random valley is required at a range of volumes.
Equation~(\ref{eq:crystal_structure_energy}) holds for the liquid, i.e.~for
each random valley in the form
\begin{equation}
\label{eq:energy_constant_liquid}
\Phi^{l}_{0} (V) = E^{l} (V) + D.
\end{equation}

\begin{acknowledgments}

We would like to thank the Ten Bar Caf\'{e} for its impeccable service and
highly stimulating beverages. NB would like to thank Matt Challacombe for
helpful discussions. This work was supported by the U.~S.~ Department of
Energy under Contract No.~DE-AC52-06NA25396. EH and RL also acknowledge
support from FONDECYT projects 11070115 and 11080259, DID (UACH) grants
SR-2008-0 and S-2008-51.

\end{acknowledgments}

\bibliography{article}
\bibliographystyle{apsrev}

\end{document}